\documentclass[preprint2]{aastex}
\usepackage[T1]{fontenc}
\usepackage{psfig}
\usepackage{latexsym}
\usepackage{mathrsfs}
\usepackage{amsmath}

\advance \voffset by 1.00cm\relax
\def\msun{{\,M_{\odot}}}

\def\mdot{\dot M}

\newcommand{\der}[2]{\ensuremath{\frac{{\rm d} #1}{{\rm d} #2}}}

\newcommand{\be}{\begin{equation}}
\newcommand{\ee}{\end{equation}}
\newcommand{\bea}{\begin{eqnarray}}
\newcommand{\eea}{\end{eqnarray}}

\makeatletter

\newenvironment{figurehere}
  {\def\@captype{figure}}
  {}
\makeatother

\begin{document}

\title{Slim disks around Kerr black holes revisited}

\author{Aleksander S\k{a}dowski}

\affil{Nicolaus Copernicus Astronomical Center,
            Bartycka 18, 00-716 Warszawa, Poland}



\begin{abstract}
We investigate
stationary slim accretion disks around Kerr black holes. We 
construct a new numerical method based on the relaxation 
technique. We systematically cover the whole parameter space
relevant to stellar mass X-ray binaries. We also notice some 
non-monotonic features in the disk structure, overlooked in 
previous studies.
\end{abstract}


\section{INTRODUCTION}

Observations of microquasars in thermal states \citep[see
e.g.][]{mcclintockremillard03}, as well as observations of some
AGN \citep[e.g.][]{collinkawaguchi04}, point to the existence of
black hole (BH) accretion flows that are quasi-steady for relatively
long times, optically thick, and geomerically thin. They exist in
a rather wide range of accretion rates.
Many authors study such flows using the celebrated Shakura-Sunyaev
model. The model is easy in applications because of its remarkable
advantages, in particular: [$1$]~Within the disk, matter rotates
on almost exact circular Keplerian orbits. [$2$]~The inner edge of
the disk, i.e. the place where the accretion flow changes from
rotation to plunging, {\it always} locates at $r_{in} = {\rm
ISCO}$. [$3$]~The local flux of radiation $F(r)$ emitted from the
disk surface is given by a universal analytic formula, derived
directly from conservation of mass, energy, and angular momentum,
and independent of dissipation. The general relativistic version
of it, $F(r) = F_{NT}(r)$, was derived by \cite{nt}.

However, for high accretion rates, e.g. ${\dot M} \ge 0.1\,{\dot M}_{Edd}$,
these properties do not hold, and the Shakura-Sunyaev
model should be changed in order to properly describe the ``slim
disk'' effects \citep{slim}: [$1^*$]~Within most of the disk the
rotation is slightly sub-Keplerian, and in the innermost part of
the disk the rotation is super-Keplerian. [$2^*$]~With the
increasing accretion rate, the inner edge of the disk $r_{in}$
goes {\it closer} to the BH than the ISCO (e.g. $\Delta r_{in}/r_{\rm ISCO} \approx
10\%$ for the Eddington mass accretion rate).  [$3^*$]~The local flux of radiation emitted from the disk
surface is not given by the universal analytic Novikov-Thorne
formula, but should be calculated to include advection of heat
captured within the accreted matter. Slim disks are far less
radiatively efficient than the Shakura-Sunyaev solutions. In this respect
they resemble advection dominated accretion flows \citep[for a recent review
see][]{narayanmcclintock08} and together with them form a class or radiatively
inefficient flows (RIFs).

Most of the RIFs models in Kerr geometry constructed to date
\citep{adafs,gammie} described advection dominated flows rather than slim disks. The
known slim disk solutions do not provide a sufficient coverage of
the parameter space needed for accurate fitting of the calculated
continuum spectra to these observed. The previous fits
\citep[e.g.][]{shafee-06} have been done with the assumption that
$r_{in} = {\rm ISCO}$ and $F(r) = F_{NT}(r)$. This is clearly
inadequate for $L > 0.3L_{Edd}$, as Figures 7 and 8
in \cite{shafee-06} show. In this paper, we have calculated a
large set of slim disk models that cover all the relevant
parameter space. Hopefully, they will improve the accuracy of the
spectral fits as well as other astrophysical applications of black
hole accretion disk models.

The paper is organized as following: At the beginning
(\S\ref{s.relativistic}) we present equations governing
relativistic accretion disk for optically thick case. In
\S\ref{sect.numerical} we describe in details the numerical
methods we use to solve the problem. The solutions for various
disk parameters are presented in \S\ref{s.results}. In
\S\ref{s.diskstructure} we put attention to non-monotonic features
in radial velocity profiles which appear for some particular mass
accretion rate range. In the following paragraph (\S\ref{s.regimes})
we discuss
properties of slim disk solutions in different regimes of the
system parameters. Finally, in \S\ref{s.conclusions} we
summarize our work and discuss astrophysical application of the
results presented here.

\section{RELATIVISTIC SLIM DISKS}
\label{s.relativistic}
\label{eq.equations}

In this section we present slim disk equations derived in the Kerr
spacetime metric. We 
follow the authors who have derived these
equations previously, starting from \cite{lasota94} and taking
into account further improvements by \cite{adafs} and \cite{vertical}.
Similar models were constructed and solved by \cite{gammie} and \cite{beloborodov98}.
 We assume: the Kerr metric, axis symmetry ($\partial_\phi=0$), 
stationarity ($\partial_t=0$), zero torque close to the horizon, large
optical depth, no self-irradiation, no angular momentum taken away by
radiation and we neglect the magnetic pressure.

(i) The mass conservation:
\begin{equation}
 \dot M=-2\pi \Sigma\Delta^{1/2}\frac{V}{\sqrt{1-V^2}}
\label{eq_cont2}
\end{equation}
where $\Sigma=\int_{-h}^{+h}\rho\,dz$ is disk surface density and $V$, defined by the relation $u^r=V\Delta^{1/2}/(r\sqrt{1-V^2})$, is the gas radial velocity as measured by an observer at fixed $r$ who corotates with the fluid. $\Delta$ and $A$ are the standard Kerr metric coefficients.

(ii) The radial momentum conservation:
\begin{equation}
\frac{V}{1-V^2}\frac{dV}{dr}=\frac{\cal A}{r}-\frac{1}{\Sigma}\frac{dP}{dr}
\label{eq_rad3}
\end{equation}
where
\begin{equation}
{\cal A}=-\frac{MA}{r^3\Delta\Omega_k^+\Omega_k^-}\frac{(\Omega-\Omega_k^+)(\Omega-\Omega_k^-)}{1-\tilde\Omega^2\tilde R^2}
\label{eq_rad4}
\end{equation}
and $\Omega=u^\phi /u^t$ is the angular velocity with respect to the stationary observer, $\tilde\Omega=\Omega-\omega$ is the angular velocity with respect to the inertial observer, $\Omega_k^\pm=\pm M^{1/2}/(r^{3/2}\pm aM^{1/2})$ are the angular frequencies of the corotating and counterrotating Keplerian orbits and $\tilde R=A/(r^2\Delta^{1/2})$ is the radius of gyration.

(iii) The angular momentum conservation \citep{lasota94}:
\begin{equation}
 \frac{\dot{M}}{2\pi}({\cal L}-{\cal L}_{in})=\frac{A^{1/2}\Delta^{1/2}\gamma}{r}\alpha P
\label{eq_ang6}
\end{equation}
where ${\cal L}=u_\phi$, ${\cal L}_{in}$ is the angular momentum at the disk inner edge, $\gamma$ is the Lorentz factor and $P=2Hp$ can be considered vertically integrated pressure. Herein, unless stated otherwise, we assume $\alpha=0.1$.

(iv) The vertical equilibrium \citep{vertical}:
\begin{equation}
 \frac{P}{\Sigma H^2}=\frac{{\cal L}^2-a^2(\epsilon^2-1)}{2 r^4}\equiv{\cal G}
\label{eq_vert}
\end{equation}
with $\epsilon=u_t$ being the conserved energy associated with the Killing vector.

(v) The energy conservation:
\begin{eqnarray}\nonumber
 &&F^{adv}\equiv-\frac{\alpha P A\gamma^2}{r^3}\frac{d\Omega}{dr}-\frac{32}{3}\frac{\sigma T^4}{\kappa\Sigma}=-\frac{\dot M}{2\pi r^2}\frac{P}{\Sigma}\times\\&&\left(\frac{4-3\beta}{\Gamma_3-1}\der{\ln T}{\ln r}-(4-3\beta)\der{\ln \Sigma}{\ln r}\right)
\label{eq.fadv2}
\end{eqnarray}
where the disk central temperature $T$ has been introduced.

(vi) The regularity condition:\\
By a series of algebraic manipulations of the above equations we get:
\be
\der{\ln V}{\ln r}=\frac{\cal N}{\cal D}(1-V^2)
\label{eq_derV}
\ee
with $\cal N$ and $\cal D$ given by:
\begin{eqnarray}\nonumber
{\cal N}&=&-{\cal A}-{\cal BC}\frac{P}{\Sigma}-\frac{4\pi r^2 F^{adv}(\Gamma_3-1)}{\dot M(1+\beta)}-\\&&\frac{1-\beta}{1+\beta}\frac P\Sigma\der{\ln\cal G}{\ln r}\\
\label{eq_derNN}
{\cal D}&=&{\cal C}\frac P\Sigma-V^2
\label{eq_derDD}
\end{eqnarray}
where: ${\cal B}$, ${\cal C}$ and $\Gamma_3$ are defined in \cite{adafs}, $\beta$ is gas to total pressure ratio and $F^{adv}$ is given by Eq. \ref{eq.fadv2}. To obtain a physical solution ${\cal N}$ and ${\cal D}$ must vanish at the same radius called \textit{the sonic point}.

\section{NUMERICAL METHODS}
\label{sect.numerical}

To obtain slim disk solutions one has to solve a two dimensional system of ordinary differential equations together with the following regularity conditions at the sonic radius $r_S$:
\be
\left.{\cal N}\right|_{r=r_S}=\left.{\cal D}\right|_{r=r_S}=0
\label{num_regcond}
\ee
and outern boundary conditions given at some large radius $r_{out}$. The location of the sonic point is not known \textit{a priori}. A mathematical problem defined in such a way can be classified as a \textit{two point boundary value problem} which we solve applying the relaxation technique \citep{numericalrecipes}. To start the relaxation process one has to provide a \textit{trial solution}. The convergence strictly depends on the quality of such initial guess. In this work we apply the following method of searching for the proper initial conditions for relaxation.

The angular momentum at the inner edge of the disk, ${\cal L}_{in}$, is the eigen value of the problem and must be chosen properly to satisfy the regularity conditions given by Eq. \ref{num_regcond}. The value of ${\cal L}_{in}$ determines the shape of a solution. It turns out that the topology of slim disk solutions with ${\cal L}_{in}$ higher and ${\cal L}_{in}$ lower than the proper value ${\cal L}_{in,0}$ is different. The former branch of solutions terminates as soon as the denominator $\cal D$ vanishes (the regularity condition is not satisfied). The latter does not cross the sonic point at all ($\cal D$ is always positive). The self-consistent solution is expected to be the common limit of these two branches. A trial solution can be achieved in a few iteration steps.\footnote{To save computational time when looking for the trial solution we estimate $\der\Omega r$ using the diffusive form of viscosity \cite[Eq. 35 in][]{adafs} instead of calculating it numerically - such an approach is precise enough to ensure convergence. This fact explains why the trial and relaxed models in Fig. \ref{f.mtrial} do not coincide.} We start integration at $r>1000r_S$ assuming the \cite{nt} boundary conditions and we use the implicit Runge-Kutta method of the 4th order.

\begin{center}
\begin{figurehere}
 \includegraphics[width=.95\columnwidth]{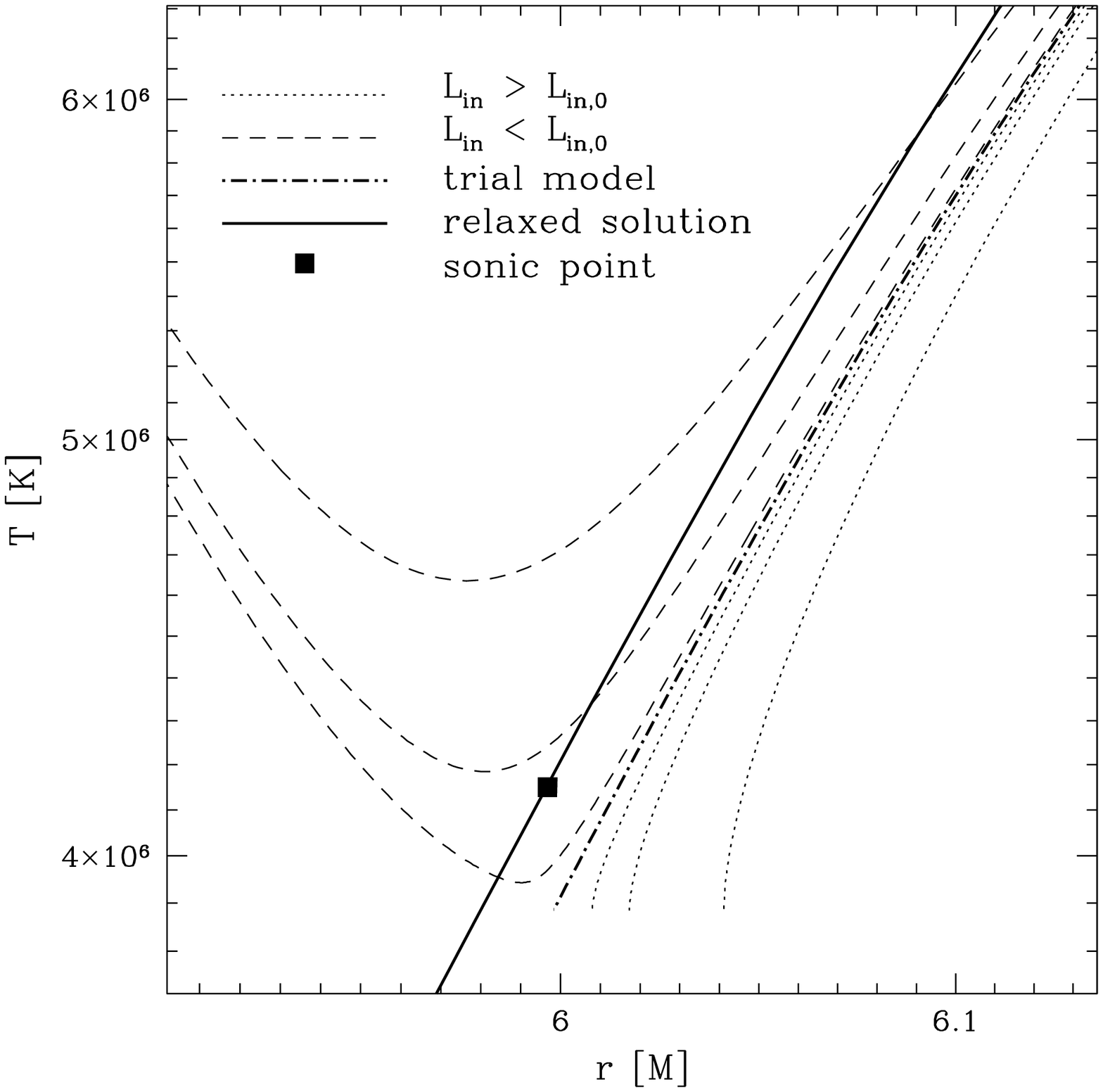}
\caption{
Temperature profiles in the vicinity of the sonic point for a few iterations leading to the trial model used as initial condition in the relaxation procedure (for details see \S\ref{sect.numerical}). According to the topology solutions with too high value of ${\cal L}_{in}$ terminate before reaching the proper sonic point while solutions with too low ${\cal L}_{in}$ go through the sonic point radius but follow an improper branch. The relaxed solution and the location of the sonic point are also presented. Models calculated for a non-spinning $9.4\msun$ BH with $\dot m=0.1\dot m_{Edd}$.
}
\label{f.mtrial}
\end{figurehere}\end{center}

Once the trial solution is found one can start relaxation process between $r_{out}$ and the estimated position of the sonic radius $r_S$. The standard approach has to be modified due to the fact that we expect singularity at the inner boundary. Thus, we treat the problem as a free boundary problem and introduce one more variable describing the position of the critical point \citep{numericalrecipes}. In this work we usually use 100 mesh points spaced logarithmically in radius.

The radial derivatives ${\rm d}\ln{\cal G}/{\rm d}\ln r$ (Eq.~\ref{eq_derNN}) and ${\rm d}\Omega/{\rm d}r$ (Eq.~\ref{eq.fadv2}) are evaluated numerically basing on the $\cal G$ and $\Omega$ profiles in the previous iteration step. A relaxed solution is obtained in a few iteration steps and is then used as an initial condition for relaxation when looking for the solution of a problem with one parameter (e.g. mass accretion rate $\dot m$ or BH spin $a^*$) slightly changed. For new system parameters we look for the outern boundary conditions at $r_{out}$ by integrating the equations from $1.1r_{out}$ (assuming Keplerian disk there) until we reach $r_{out}$. In such a way a full spectrum of system parameters can be achieved effectively.

Once a solution outside the sonic point is found we numerically estimate the radial derivatives of $V$ and $T$ at the sonic point using values given at $r>r_S$. Taking them into account we make a small step from the innermost mesh point (which corresponds to the sonic point) inward. Then we start integrating using standard Runge-Kutta method until we get close enough to the horizon.

\section{RESULTS}
\label{s.results}

\subsection{Flux profiles}
\label{s.flux}

For very low mass accretion rates most of energy generated at each radius is immediately emitted away and disk is expected to be very thin. Therefore, solutions in that regime are in general consistent with solutions of \cite{nt}. The slim disk solutions deviate from thin disk models when advection becomes important. According to Fig. \ref{f.fadvfac} in case of a non-rotating BH an accretion disk is no longer radiatively efficient for mass accretion rates exceeding $0.1\dot m_{Edd}$ where $\dot m_{Edd}$ is the critical mass accretion rate defined as:
\be
\dot m_{Edd}=\frac{64\pi GM}{c\kappa_{es}}=2.23\times 10^{18} \frac M\msun \quad\rm g\cdot s^{-1}
\label{res.mdotc}
\ee
corresponding to (in case of a non-rotating BH) a disk with the Eddington luminosity (see \S\ref{s.efficiency}). The rate of heat advected increases with mass accretion rate. One can expect that for very high mass accretion rates advection would dominate heat transfer \citep{slim}\footnote{There is another class of accretion disks which are advection dominated and opticaly thin named \textit{Advection Dominated Accretion Flows} investigated by a number of groups e.g. \cite{ichimaru77}, \cite{rees82}, \cite{narayanadafs}, \cite{narayanyi95} and \cite{thermalequilibria}}. It is important to note that inside some particular radius (e.g. $8M$ for $\dot m=0.6\dot m_{Edd}$) heat is no longer accumulated in the accreted matter but is gradually radiated away amplifying the emission coming out of viscous processes. This fact has strong influence on the radial profiles of the flux emitted from both sides of an accretion disks (Fig. \ref{f.fluxes}). For mass accretion rates implying significant amount of advection the maximum of emission is shifted inward: from $\sim10M$ expected for radiatively efficient disks around non-spinning BH down to $~7.5M$ for $\dot m=0.9\dot m_{Edd}$ and even further for higher accretion rates. The effect is even more distinct for rotating BHs (middle and bottom panels of Fig. \ref{f.fluxes}, please note the common shift due to decreasing radius of the innermost stable circular orbit).

\begin{center}
\begin{figurehere}
 \includegraphics[width=.99\columnwidth]{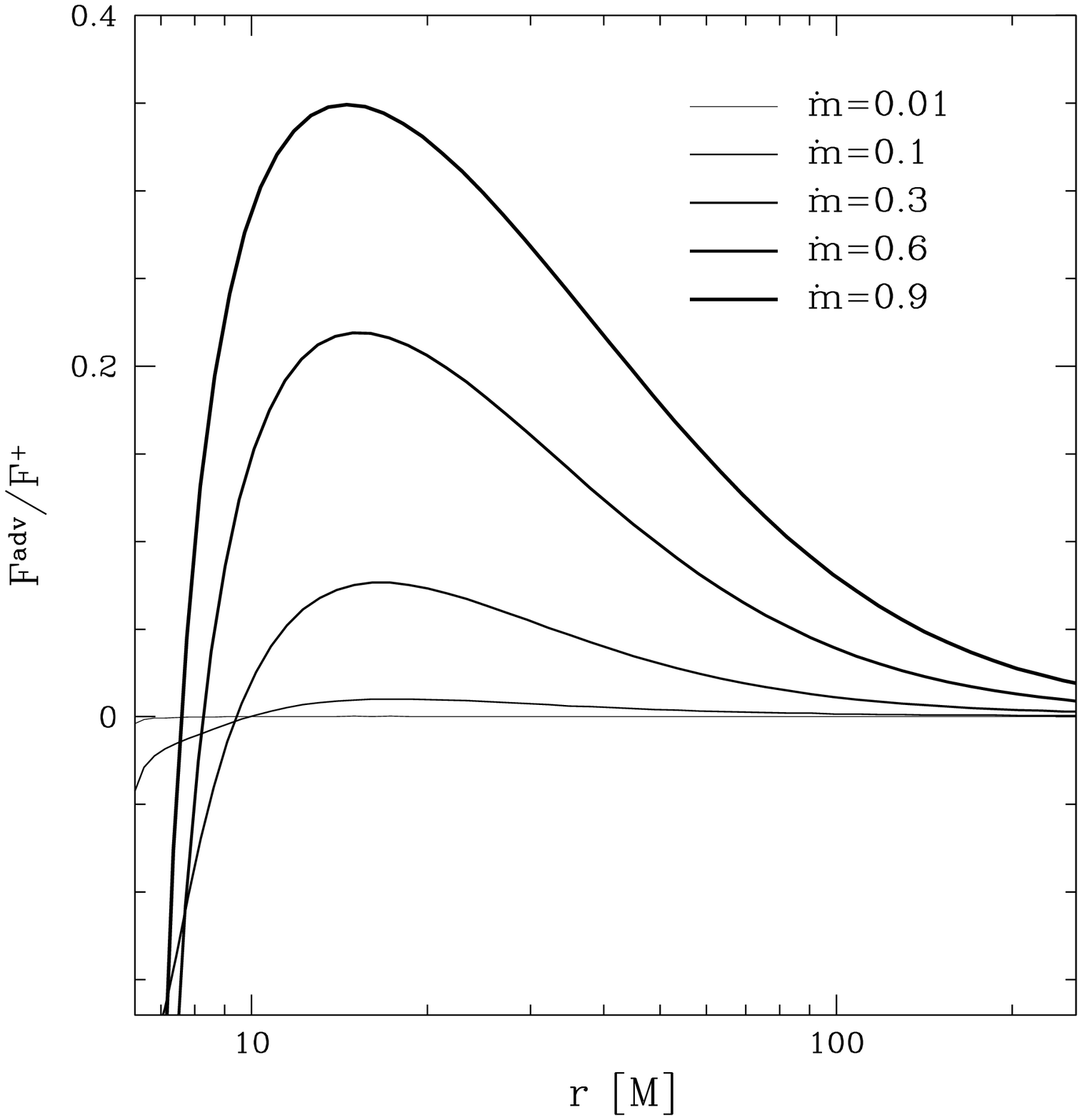}
\caption{
Ratio of energy advected to generated at each radius for different mass accretion rates for $a^*=0$. Positive values denote region where heat generated by viscous processes is stored in the accreted matter. Negative values mark radii where the cumulated heat is radiated away summing up to the flux generated by viscous processes. 
}
\label{f.fadvfac}
\end{figurehere}\end{center}

\begin{center}
\begin{figurehere}
 \includegraphics[width=.99\columnwidth]{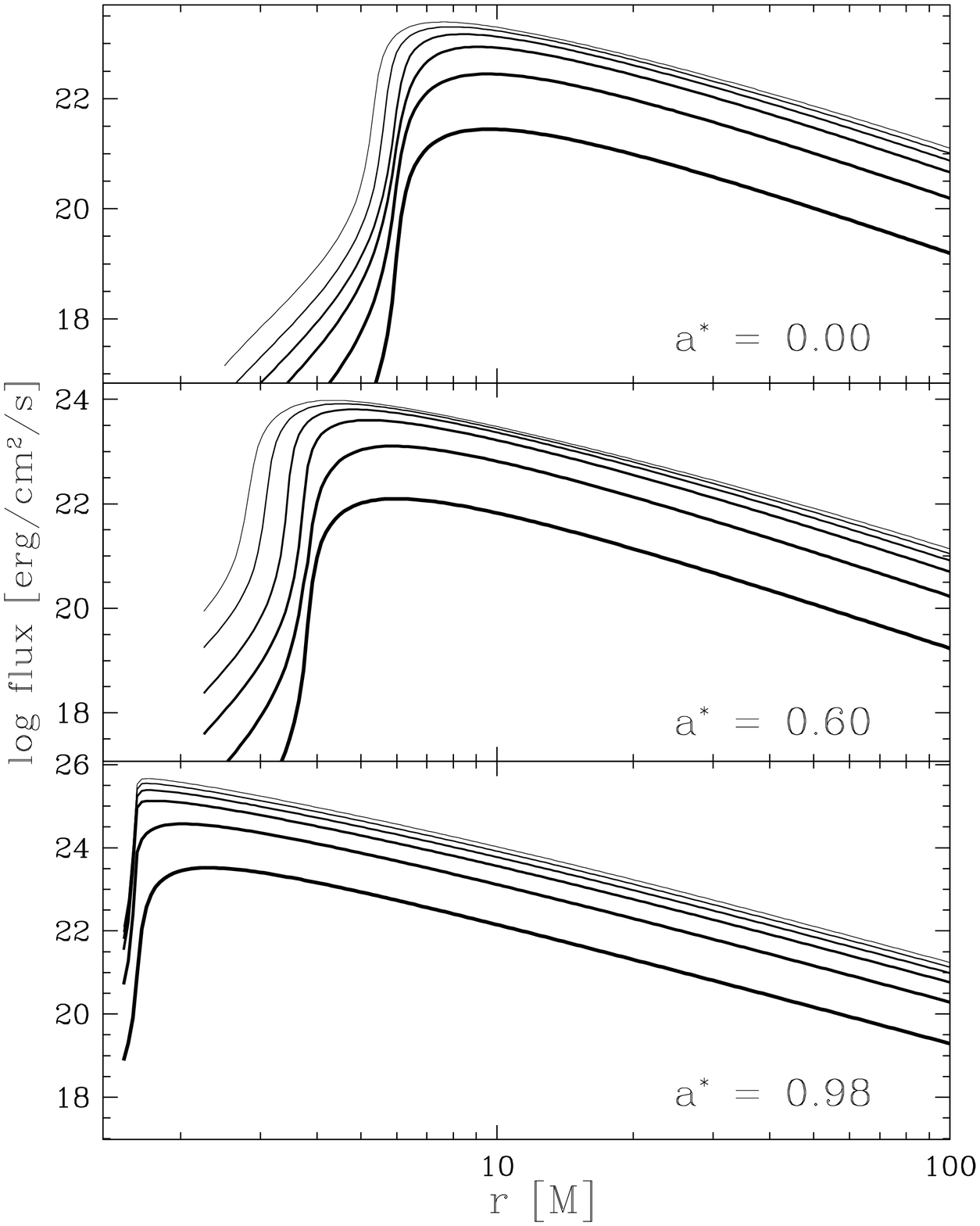}
\caption{
Flux profiles for different mass accretion rates and BH spins. Each subplot contains six solid lines for the following mass accretion rates: $0.01$ (the thickest line), $0.1$, $0.3$, $0.6$ and $0.9 \dot M_{Edd}$ (the thinnest line). The upper panel is for a non rotating BH ($a^*=0$), the middle one for $a*=0.6$ while the bottom one for a highly spinning BH ($a^*=0.98$). BH mass is $9.4\msun$.
}
\label{f.fluxes}
\end{figurehere}\end{center}

\subsection{Disk structure}
\label{s.diskstructure}

Profiles of a few parameters describing an accretion disk for a number of mass accretion rates in case of a non-spinning BH are presented in Fig. \ref{f.4plot}. The profiles of the radial velocity as measured in the fluid corotating frame $V$ are presented in the upper left panel. The ratio of the radiation to gas components of the total vertically integrated pressure $P$ are drawn in the upper right corner. Disk thickness and central temperatures are presented in the lower panels. 

The radial velocity for large radii coincide with values given by Novikov \& Thorne solutions which are assumed as the outer boundary conditions and approaches the speed of light when getting close to the horizon. At a given radius the radial velocity increases with mass accretion rate. For a specific range of accretion rates the radial velocity profiles are not everywhere monotonic i.e. the velocity is not steadily increasing when gas is getting closer to the horizon. These features, existence of which is out of common knowledge, correspond to regions of increased surface density and will be discussed in details in the following paragraph.

\begin{center}
\begin{figurehere}
 \includegraphics[width=.99\columnwidth]{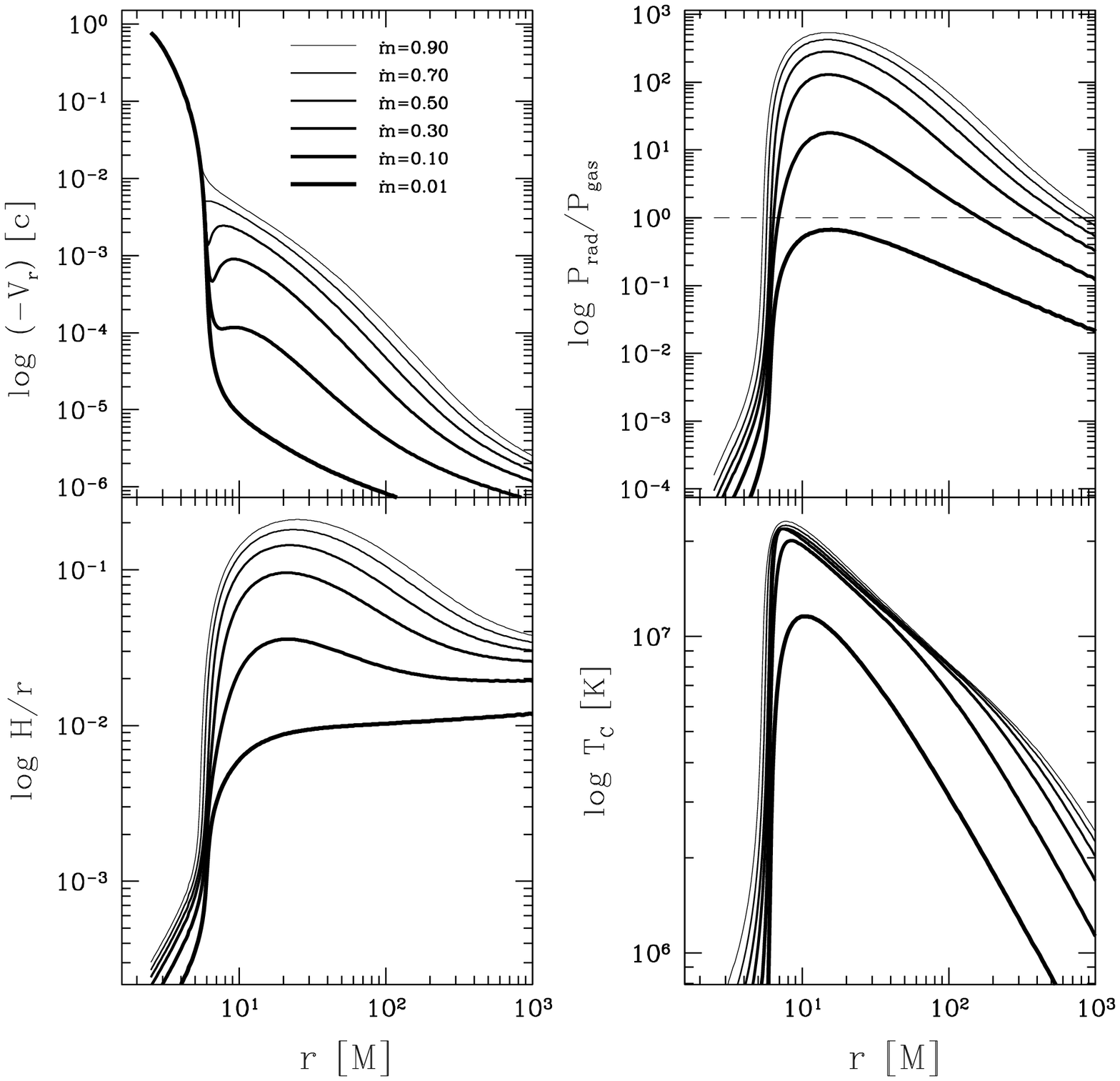}
\caption{
Profiles of radial velocity (upper-left panel), radiation to gas pressure ratio (upper-right), disk thickness over radius ($H/r$) ratio (bottom-left) and central temperature (bottom-right) for a nonspinning BH of mass $9.4\msun$. The solid lines are for mass accretion rates between $0.01$ and $0.9 \dot m_{Edd}$. 
}
\label{f.4plot}
\end{figurehere}\end{center}

\cite{shakura-73} pointed out that an accretion disk can be divided into three distinctive regimes: (i) \textit{the outer region}, gas pressure and free-free scatterings dominated; (ii) \textit{the middle region}, gas pressure but electron scattering dominated; (iii) \textit{the inner region}, radiation pressure and electron scattering dominated. According to their \textit{inner region} formulae the surface density is expected to rise infinitely when approaching the inner edge of the disk (accreted gas is slowed down by increasing radiation pressure). Obviously, this behaviour (which corresponds to a decrease of the fluid radial velocity) is suppressed as the Shakura \& Sunyaev's inner region does not infact extend down to the inner edge of the disk. The radiation to gas pressure ratio in their solutions is proportional to \citep{shakura-73}:
\be
\frac{p_{rad}}{p_{gas}}\propto r^{-8/21}\left(1-\sqrt{\frac 6r}\right)^2
\label{res.pradpgas}
\ee
The transition between \textit{inner} and \textit{middle} regions takes place where gas and radiation components of pressure are equal. It is trivial to show that the inner region must be followed by a transition to the middle one before reaching the disk inner boundary. Therefore, close to $6M$ the growth of the surface density is suppressed by a decrease expected in the gas pressure dominated (\textit{middle}) region of an accretion disk. The resulting humps in the surface density as well as the corresponding non-monotonic parts of the radial velocity profiles are presented in details in Fig. \ref{f.humps}. According to the upper-right panel of Fig. \ref{f.4plot} for the lowest mass accretion rates ($\dot m<0.01\dot m_{Edd}$) radiation pressure never dominates in the disk and therefore there is no hump in surface density profile at all. For moderate mass accretion rates radiation pressure exceeds gas pressure in wide range of radii and resulting humps in surface density profiles are clearly visible. When accretion rate increases the advection of heat becomes more and more important (see \S\ref{s.flux}) and the Shakura \& Sunyaev formalism cannot be longer applied. Our solutions show that for mass accretion rates higher than $0.9\dot m_{Edd}$ (for $a^*=0$) the surface density and radial velocity profiles become again monotonic for all radii despite the fact that inner parts of the disk are radiation pressure dominated. It is important to understand that such humps in surface density profiles cannot be considered shock like features - they are perfectly continuous, finite and stationary. Their shape depends on the assumptions we make about disk vertical structure. It is possible that for more realistic models they would be even more profound. 

\begin{center}
\begin{figurehere}
 \includegraphics[width=.99\columnwidth]{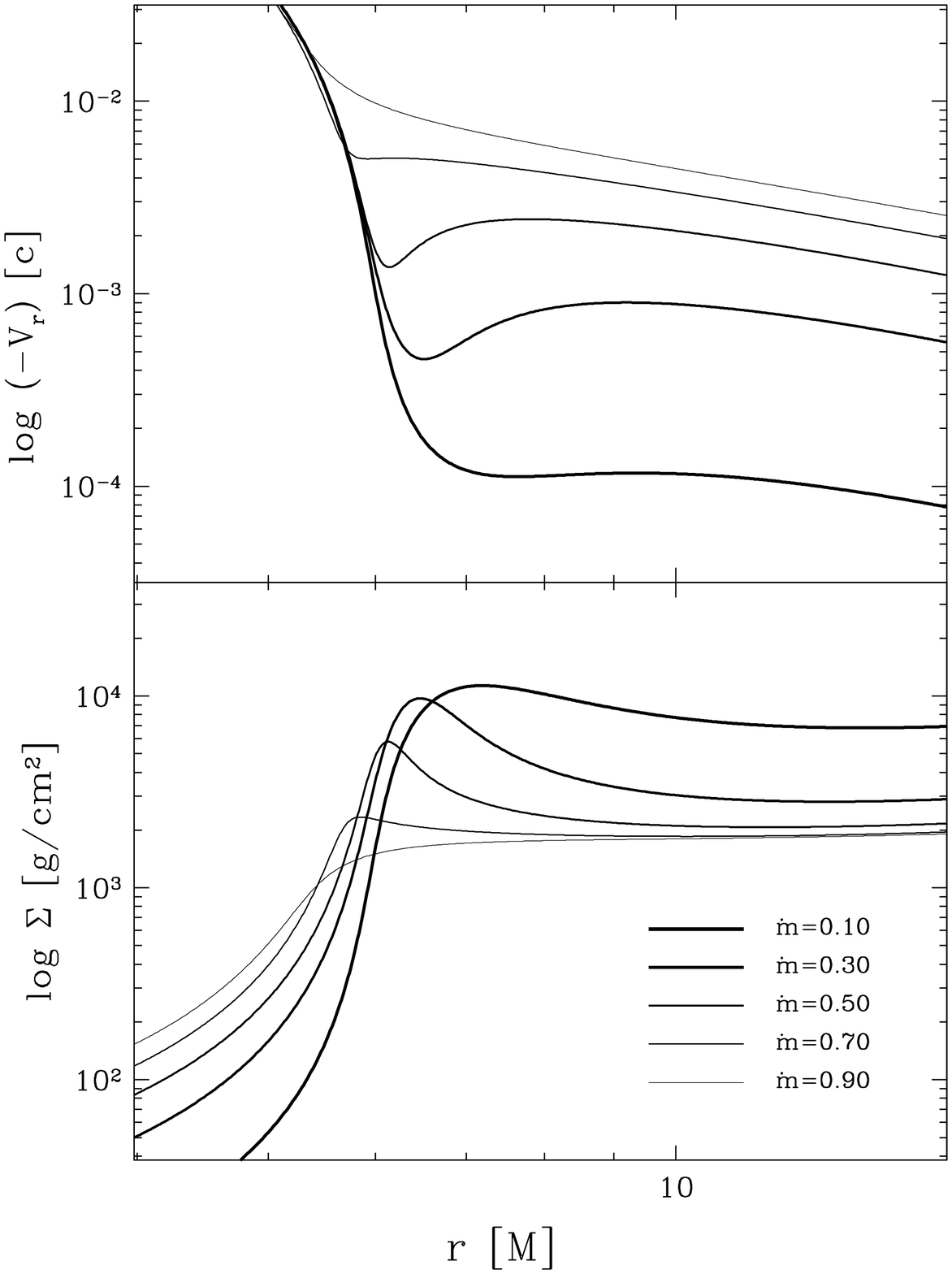}
\caption{
The non-monotonic sections of the radial velocity and surface density profiles in an accretion disk surrounding a $9.4\msun$ BH with $\alpha=0.1$. The solid lines are for mass accretion rates between $0.1$ and $0.9 \dot m_{Edd}$.
}
\label{f.humps}
\end{figurehere}\end{center}

The local minima in the radial velocity profiles occur in some particular range of mass accretion rates. The minimal accretion rates for which the non-monotonic features occur correspond to the appearance of radiation pressure dominated disk regions. The maximal values are determined by the rate of advection which significantly changes the disk structure. Fig. \ref{f.dumps} presents relation between the mass accretion rates for which we observe non-monotonic sections in the radial velocity profiles and the BH spin. For non-rotating BH the local minima appear for accretion rates between $0.1$ and $0.7\dot m_{Edd}$ while for a highly spinning BH ($a^*=0.9$) for accretion rates range $0.025\div0.2\dot m_{Edd}$. The solid line presents mass accretion rates for which the minimum is the deepest (compare Fig. \ref{f.humps}). In general the minima are most profound for mass accretion rates lying almost perfectly in between the limiting values i.e. $0.38$ for $a^*=0$ and $0.11$ for $a^*=0.9$. These deepest minima occur at radii plotted in the bottom panel of Fig. \ref{f.dumps}. They lie just outside the innermost stable orbit for a given BH spin.

\begin{center}
\begin{figurehere}
 \includegraphics[width=.99\columnwidth]{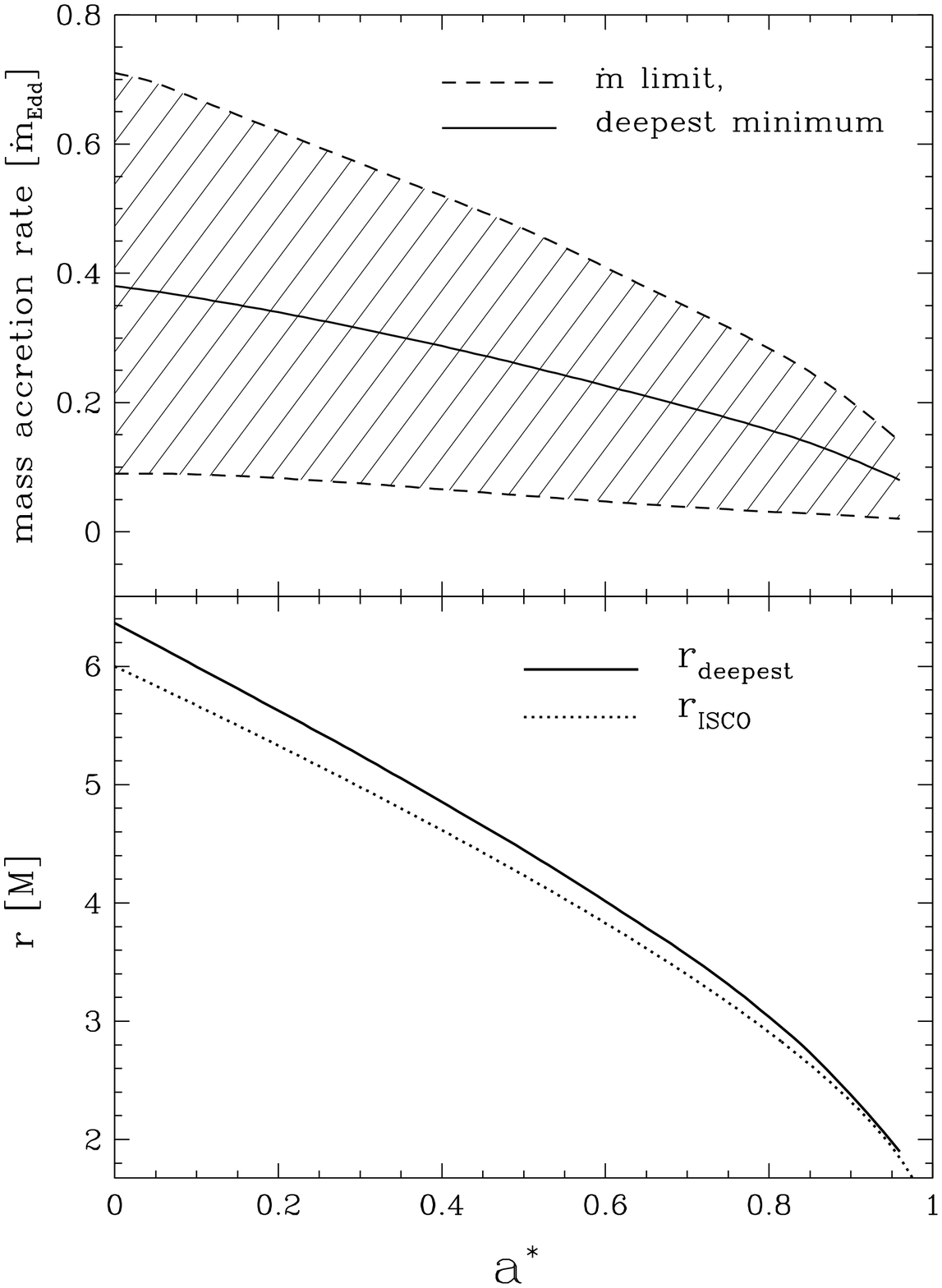}
\caption{
The shaded area at the upper panel presents mass accretions rates for which the non-monotonic sections of the radial velocity profile (see \S\ref{s.diskstructure}) appear at a given BH spin. These features are the most profound at the mass accretion rate given by the solid line. The bottom panel presents location of the most profound local minimum in the radial velocity profile as a function of BH spin. Location of the innermost stable orbit is also presented.
}
\label{f.dumps}
\end{figurehere}\end{center}

The bottom left panel of Fig. \ref{f.4plot} presents the $H/r$ ratio for a few values of the accretion rate. For disks with the lowest $\dot m$, which are gas pressure dominated only, the $H/r$ ratio is almost constant down to $\sim10M$ and then rapidly decreases. For higher accretion rates there is a radiation pressure dominated inner region of a disk which results in an increase of $H/r$ ratio.

The central temperature profile dependence on the mass accretion rate is presented in the bottom right panel of Fig.~\ref{f.4plot}. The exponents of the temperature profiles outside $\sim 10M$ depend on the gas to radiation pressure ratio. The temperature in the gas dominated regions increase more rapidly with decreasing radius than in the inner radiation pressure dominated part of a disk. As in the case of the flux profiles (Fig.~\ref{f.fluxes}) the location of the temperature maximum moves closer to the ISCO with increasing mass accretion rate. In the plunging region the temperature drops rapidly and the disk becomes gas pressure dominated.

\subsection{Angular momentum profiles}
\label{s.inneredge}
In Fig. \ref{f.ell} we present profiles of the specific angular momentum $\ell=u_\phi/u_t$ of the accreted gas for few different mass accretion rates in case of a non- and highly spinning BHs. For low and moderate mass accretion rates the specific angular momentum profiles are very close to the Keplerian distribution. They differ significantly for high mass accretion rates. As one could expect for an accretion disk \citep{scholarpedia} the flow is sub-Keplerian at large radii (the higher mass accretion rate the more significant deviation from the Keplerian flow). There is a transition radius ($r_{center}$) where the specific angular momentum crosses the Keplerian profile. This radius corresponds to the location of the pressure maximum. Inside $r_{center}$ the angular momentum is superkeplerian and crosses again the Keplerian profile at $r_{in}$ interpreted as the disk inner edge. The specific angular momentum profiles approach the values corresponding to the angular momentum at the inner edge of a disk (${\cal L}_{in}$).

\begin{center}
\begin{figurehere}
 \includegraphics[width=.99\columnwidth]{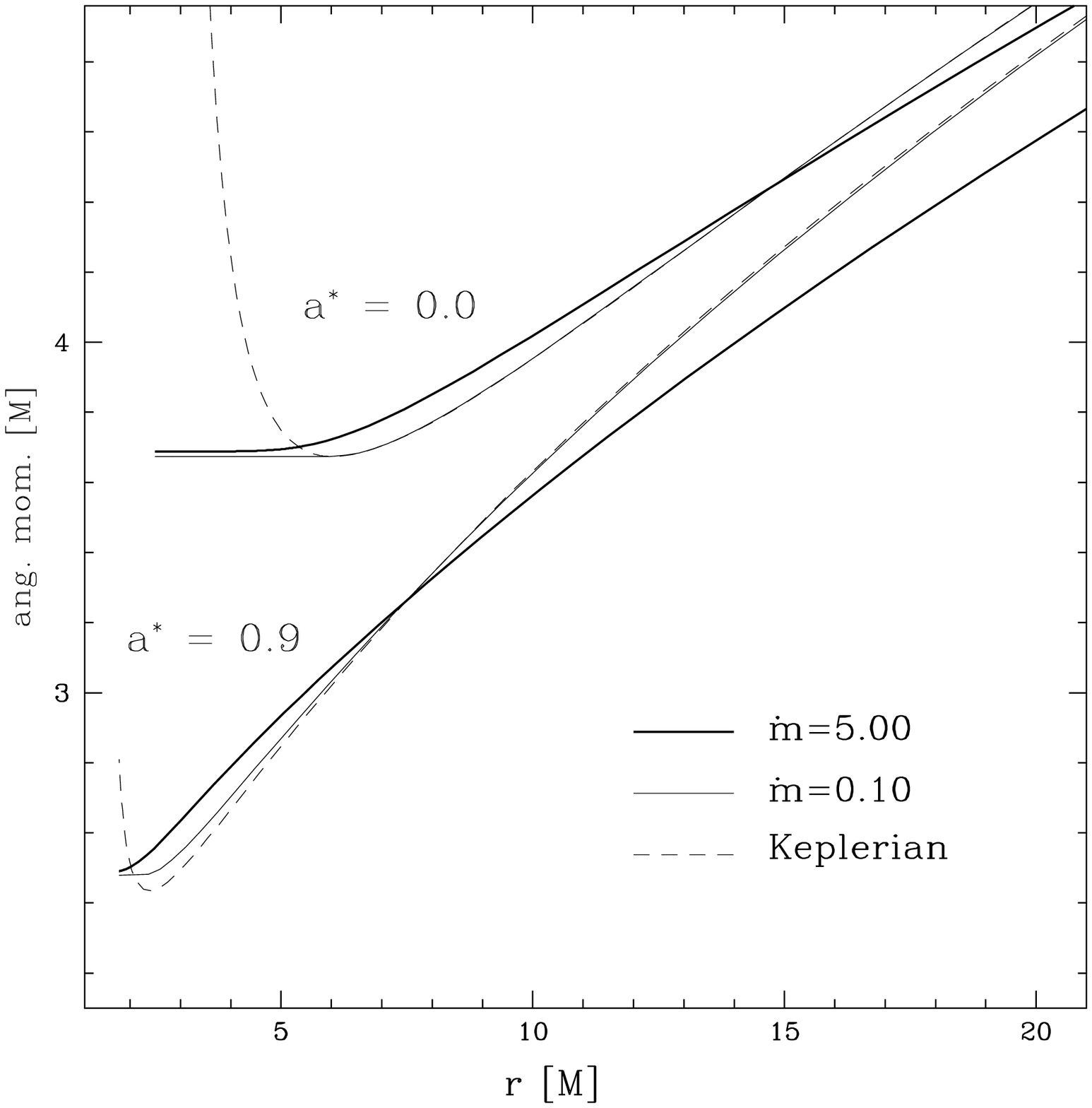}
\caption{
Accretion disk specific angular momentum profiles for non-spinning ($a^*=0.0$) and rapidly rotating ($a^*=0.9$) BHs. For each value profiles for two mass accretion rates are presented: $\dot m = 0.1$ and $\dot m = 5.0 \dot m_{Edd}$.
}
\label{f.ell}
\end{figurehere}\end{center}

In Fig. \ref{f.rinrc} we present the dependence of the location of disk characteristic radii ($r_{in}$, $r_S$ and $r_{center}$) on the mass accretion rate for two values of BH angular momentum. For low mass accretion rates ($\dot m<0.6\dot m_{Edd}$ for $a^*=0$ and $\dot m<0.2\dot m_{Edd}$ for $a^*=0.9$) the location of the disk inner edge ($r_{in}$) coincides with the location of the sonic point ($r_{S}$) and is almost independent of the accretion rate. In this regime the disk inner edge is located very close to the ISCO (marked with dotted lines). For higher mass accretion rates both $r_{in}$ and $r_{S}$ move closer to the horizon. Initially, the sonic point moves inward slower and it does not coincide with the inner edge of the disk. For the highest mass accretion rates they are again located close to each other. The location of the pressure maximum ($r_{center}$) is roughly independent of the accretion rate for low accretion rates. For $\dot m>0.6\dot m_{Edd}$ in case of a non-rotating BH and $\dot m>0.2\dot m_{Edd}$ for $a^*=0.9$ it moves slightly inward reaching $\sim10M$ for a non-rotating and $\sim4M$ for a highly-spinning BH at $\dot m=10\dot m_{Edd}$.

A more detailed discussion of several astrophysically
interesting issues connected to the location of the
inner edge of slim disks, that follow from the
transonic models calculated here, will be soon
published elsewhere.

\begin{center}
\begin{figurehere}
 \includegraphics[width=.99\columnwidth]{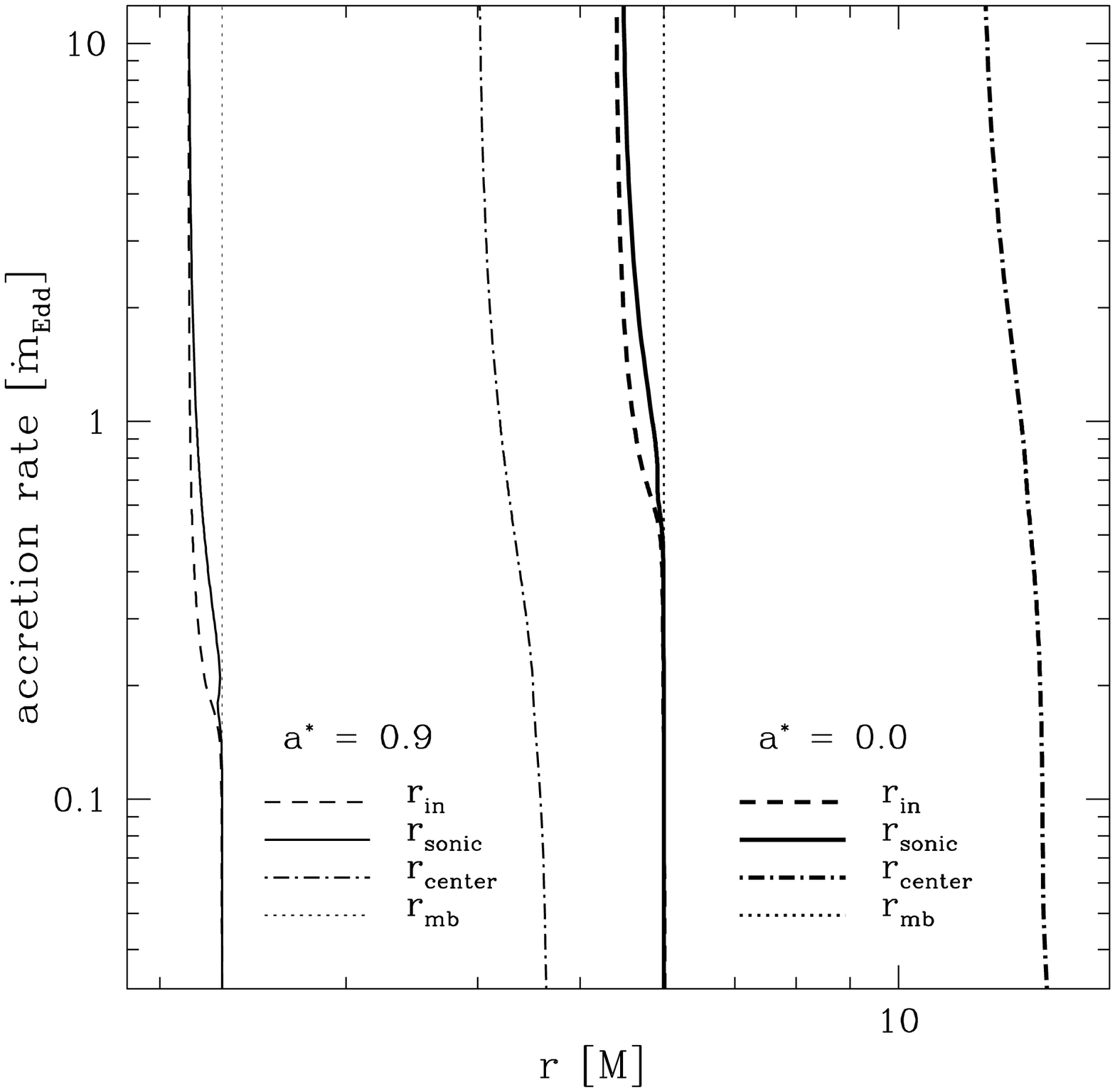}
\caption{
Locations of the characteristic points of a disk for non-spinning ($a^*=0.0$, thick lines) and rapidly rotating ($a^*=0.9$, thin lines) BHs for $\alpha=0.1$. Dashed lines mark location of the inner edge of the disk, solid lines stand for the sonic point while dot-dashed lines are for the location of the pressure maximum. Dotted lines denote location of the innermost stable circular orbits.
}
\label{f.rinrc}
\end{figurehere}\end{center}

\subsection{Radiative efficiency}
\label{s.efficiency}


In Fig.~\ref{f.eff} we plot the radiative efficiency of accretion,
defined as:
\be 
\label{res.efffactor} \eta=\frac{L}{\dot m c^2}=\frac 1{16}\frac{L/L_{Edd}}{\dot m/\dot m_{Edd}}
\ee 
versus mass accretion rate for a few values of the BH spin.
As discussed by e.g. \cite{kozlowski78} or \cite{jaroszynski80} for 
accretion disks that are thin on their inner edge, the efficiency 
of accretion $\eta$ can be approximated by $\eta = 1 - u_t(r_{in})$.
As it was discussed in the previous paragraph, when the accretion rate 
is very small, the inner edge of the disk coincides with the location 
of the ISCO (Fig. \ref{f.rinrc}). Thus, for very small accretion rates, the
efficiency is constant (does not depend on the accretion rate),
and the total luminosity is proportional to the accretion rate.

It was found by \cite{jaroszynski80}, \cite{slim}, \cite{paczynski1982}, \cite{paczynski1998} and several other authors, that when
accretion rate increases, the inner edge of a disk moves from
the location of ISCO to the location of the marginally bound orbit
$r_{mb}$ (cf. Fig. \ref{f.rinrc}). Because $u_t(r_{mb}) = 1$, the
efficiency at $r_{MB}$ is formally zero, and therefore, as found
in the above quoted papers, when accretion rate increases,
luminosity is not proportional to the accretion rate, but grows
more slowly. We illustrate the
drop in efficiency in Fig. \ref{f.eff}.

A more detailed discussion of the efficiency of accretion at high
accretion rates may be found in the review by \cite{abramowicz2005}.


\begin{center}
\begin{figurehere}
 \includegraphics[width=.99\columnwidth]{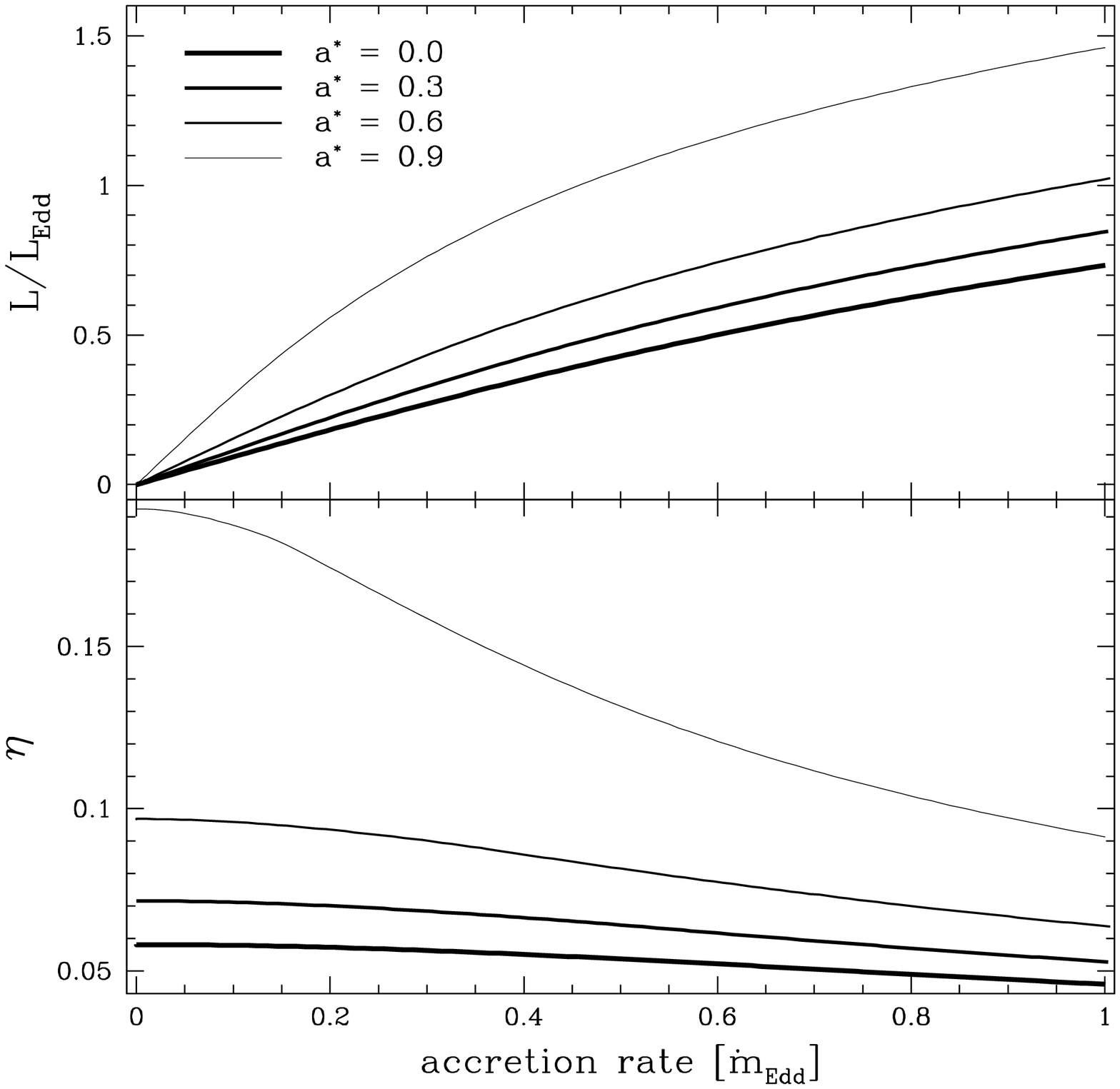}
\caption{
The top panel present the total luminosity of an accretion disk for different values of BH spin as a function of mass accretion rate. The bottom panel presents the efficiency parameter $\eta$ (see \S\ref{s.efficiency}).
}
\label{f.eff}
\end{figurehere}\end{center}

\subsection{Alpha dependence}
\label{s.alphas}

The $\alpha$ formalism \citep{shakura-73} gives only the upper limit for the
value of $\alpha$ parameter which has to fullfil the condition $\alpha\le1$.
This restriction comes from the assumption that turbulent elements of size
smaller that disk thickness and moving at velocities smaller than speed of sound
waves are the source of viscosity. The usual approach is to set $\alpha$
to some particular constant value. In this paper we choose $\alpha=0.1$ as our standard assumption.

Fortunately, as \cite{shakura-73} have proven, the outcoming flux of energy does 
not depend on $\alpha$ in the case of radiatively efficient accretion disks.
Whether this is true or not for disks with advection has to be checked numerically.
In Fig. \ref{f.alphas} we present our solutions for a few different values of the $\alpha$
parameter and for two mass accretion rates. The surface density profiles depend strongly on $\alpha$ - to generate the same amount of energy a model with lower value of $\alpha$ requires higher value of the vertically integrated pressure which corresponds to higher column density. According to the bottom panels of Fig. \ref{f.alphas} the flux profiles are insensitive to different values of $\alpha$ at all radii outside the ISCO even for high mass accretion rates. The differences in flux profiles inside the ISCO do not influence the emergent spectra as the emission in this region is negligible.

\begin{center}
\begin{figurehere}
 \includegraphics[width=.99\columnwidth]{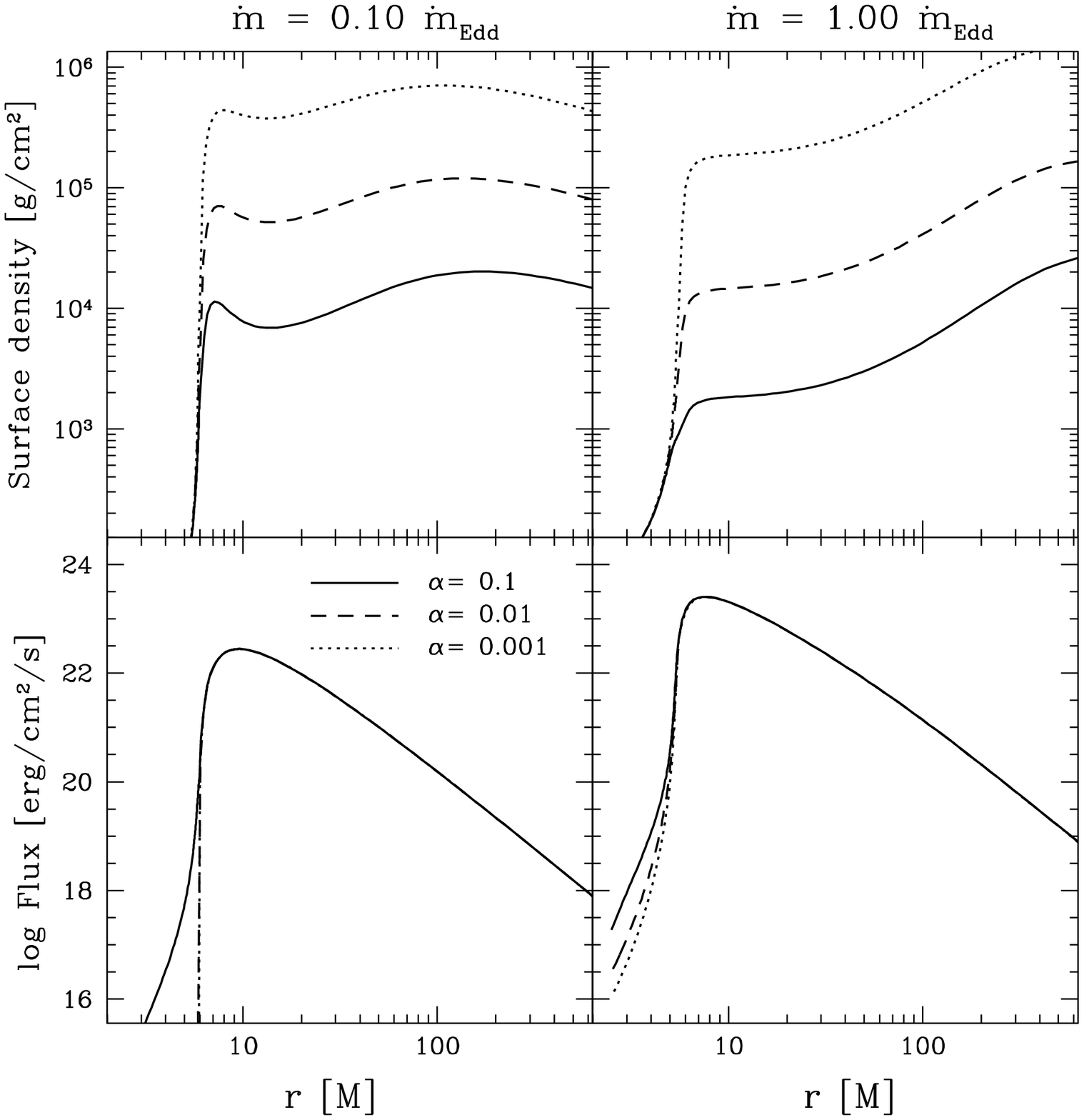}
\caption{
Surface density and flux profiles dependence on the value of the $\alpha$ parameter for
two mass accretion rates $\dot m=0.10\dot m_{Edd}$ (left panels) and $1.00\dot m_{Edd}$
(right panels). BH mass is $9.4\msun$.
}
\label{f.alphas}
\end{figurehere}\end{center}

In Fig. \ref{f.rsonic} we present the location of the sonic point for different values of the  $\alpha$ parameter and different mass accretion rates for a non-rotating BH. For very low accretion rates the sonic point is located close to the ISCO independently of the value of the visosity parameter. For high mass accretion rates and low viscosities the sonic point, as the theory of thick accretion disk predicts \citep[e.g.][]{jaroszynski80,paczynski1982}, moves inward towards the marginally bound orbit. For the highest values of the $\alpha$ parameter ($\alpha\gtrsim0.2$) the behaviour is opposite - the sonic point moves outside the ISCO with increasing mass accretion rates \citep[compare Fig.\ 11 in][]{slim}. Such solutions are interpreted as Bondi-like and are not unique \citep{muchotrzebczerny86}. Therefore the dotted lines in Fig.~\ref{f.rsonic} should be interpreted as profiles describing approximate mean locations of the sonic point in this regime.

\begin{center}
\begin{figurehere}
 \includegraphics[width=.99\columnwidth]{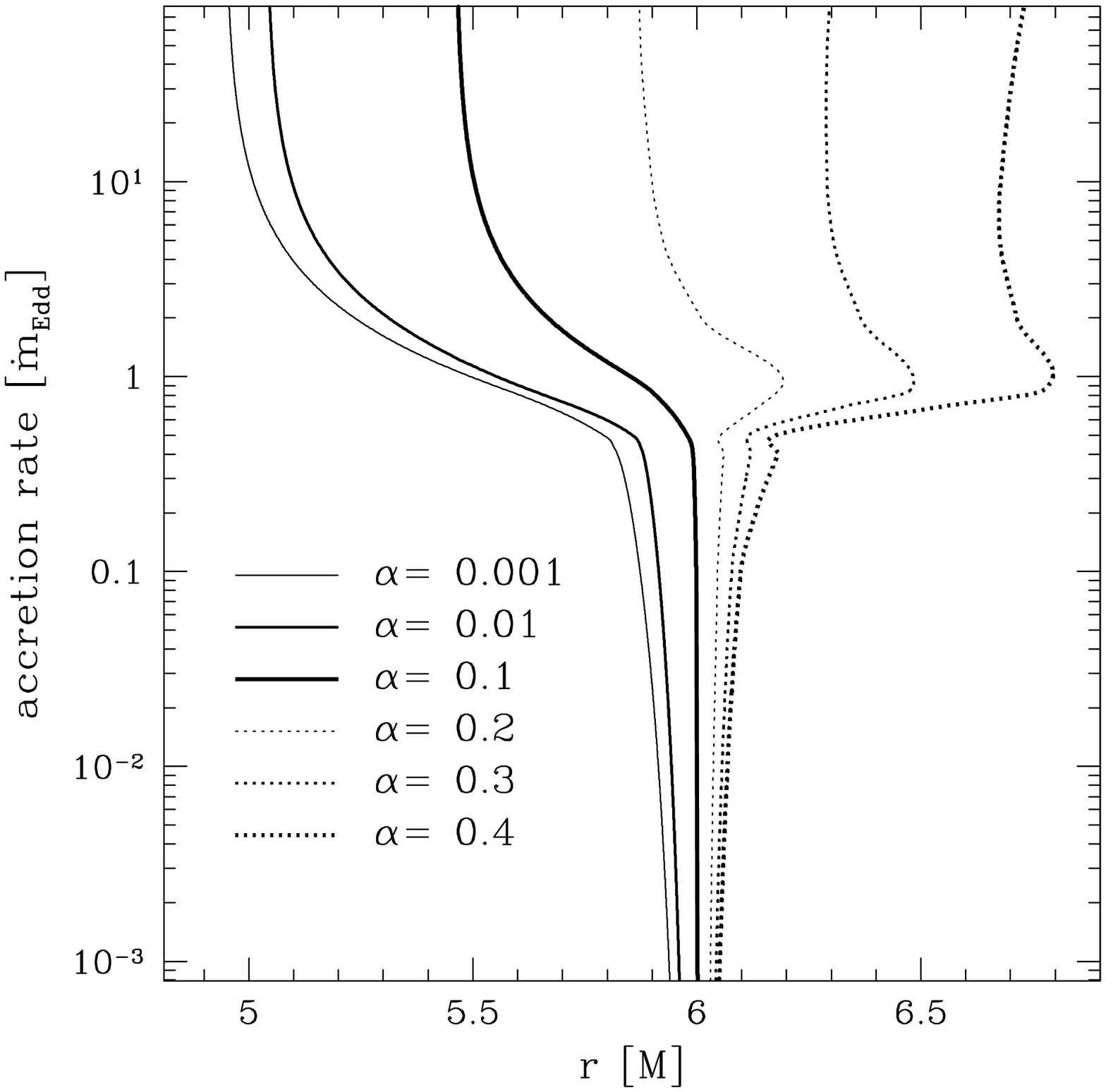}
\caption{
Location of the sonic point as a function of mass accretion rate for different values of the $\alpha$ parameter: $\alpha = 0.001$, $0.01$, $0.1$, $0.2$, $0.3$, and $0.4$ for a non-rotating BH. Dotted lines (for highest values of $\alpha$) denote approximate mean location of the sonic point as the solutions in this regime are not unique (see \S\ref{s.alphas})
}
\label{f.rsonic}
\end{figurehere}\end{center}

\section{Parameter study}
\label{s.regimes}

The main goal of this work is to provide a spectrum of slim disk solutions in a wide range of parameters applicable to stellar mass X-ray binaries. The parameter space is spanned by four values defining the system properites: BH mass, accretion rate, BH angular momentum and the value of $\alpha$. In the standard Novikov \& Thorne approach the flux profile (most important for spectral continuum fitting) does not depend on the value of $\alpha$ and is inversely proportional (for given $r/M$ and $\dot m/\dot m_{Edd}$) to the BH mass. Our results show that taking advection into account does not change these relations, e.g. the flux profile hardly depends on $\alpha$ (\S\ref{s.alphas}) neither the advection rate (cf.~Fig.~\ref{f.fadvfac}) does depend on the BH mass. Therefore, the major properties of an accretion disk depend mostly on two parameters: the mass accretion rate $\dot m$ and the dimensionless spin parameter $a^*=a/M$. In this section we point out the existence and summarize the properties of three regimes in $\dot m$ corresponding to three different branches of the $M(\Sigma)$ curve described in details in \cite{slim}. The mass accretion rates limiting them depend on BH spin and coincide with accretion rate limits for the non-monotonic features in the radial velocity profiles presented in the top pannel of Fig.~\ref{f.dumps}.

(i) \textit{low accretion rates} --- Disk is gas dominated only. The radiation pressure supported inner region does not occur. The radial velocity profile is monotonic. Disk is radiatively efficient ($\eta\approx1$ for $a^*=0$ and $\eta\gg1$ for $a^*\rightarrow1$). The advection is not significant. The flux profile agrees with the Novikov \& Thorne solution. The angular momentum profile is almost Keplerian. Both $r_{in}$ and $r_S$ are very close to the ISCO.

(ii) \textit{moderate accretion rates} --- Radiation pressure dominates the inner region of a disk. The radial velocity and surface density profiles are no longer monotonic. Disk is not radiatively efficient: $\eta$ drops down with increasing accretion rate. The advection becomes more significant what results in a shift of the flux profile maximum with respect to the radiatively efficient case. Rotation is slightly sub-Keplerian outside $r_{center}$ and super-Keplerian between $r_{center}$ and $r_{in}$. $r_{in}$ and $r_S$ almost coincide and are located close to the ISCO.

(iii) \textit{high accretion rates} --- The radiation pressure dominated region extends upto large radii. The advection becomes dominant significantly shifting the flux maximum inward. The radial velocity profile is monotonic. The angular momentum profile significantly deviates from the Keplerian distribution. Disk is radiatively inefficient ($\eta\rightarrow0$). $r_{in}$ and $r_S$ are located inside the ISCO.

\section{Prospects for future work}
\label{s.prospects}

Models of slim accretion disks calculated here are
needed in several astrophysical applications that are
at present being worked by us in a framework of a
larger research project (connected to my Ph.D. work)
that includes:

 (1) Combining the transonic radial description of
 the slim disk structure used here with a more
 accurate treatment of the vertical structure by
 using ideas and methods developed to solve the
 vertical radiation transfer in thin accretion disks
 by \cite{hameury98,davisomer05,rozanskamadej08}
 and others. We are working on a general
 numerical code that would self-consistently combine
 radial and vertical structure calculations. The
 major issue here is to calculate the location of
 the effective photosphere of a slim disk. We have
 already tested a simplified version of the code and
 examined a thin, radiation pressure supported disk,
 confirming that in this case location of the
 effective photosphere only weakly depends on the
 (unknown and therefore \textit{ad hoc} assumed) vertical
 dissipation \citep{photosphere}.

 (2) Calculating the ``observed'' X-ray continuum spectra of
 slim disks by ray-tracing photon trajectories
 (in the Kerr geometry) all the way from an emission
 place at the disk effective photosphere to a
 distant observer. We use a particular version
 of the ray-tracing code developed by \cite{bursa.raytracing}.
 It fully includes all special and general
 relativistic effects. An extensive catalog of
 the calculated spectra will be published elsewhere
 \citep{spectracatalogue} and used in collaboration
 with the Harvard group \citep{spinestimates} to improve the black hole spin estimates by
 the spectral X-ray fitting \citep[as in e.g.][]{shafee-06,donegrs1915,lmcx1spin}.
 Improvements follow from including effects of
 advection (they are relevant for higher accretion
 rates) and from a more accurate calculation of
 the location of the effective photosphere (relevant
 at higher inclinations).

 (3) The stationary slim disk models calculated here
 provide the initial conditions needed in numerical
 simulations of non-stationary slim disks in Kerr
 geometry that we perform in collaboration with the
 Xiamen University group. The interesting astrophysical issue
 here is a possible limit-cycle behavior. Also in
 this case we are calculating the observed appearance
 of the disk using the ray-tracing from the effective
 photosphere.

\section{Conclusions}
\label{s.conclusions}

In this paper we presented a numerical method used to solve equations describing slim accretion disks. The most important assumptions we made were as following: we assumed stationarity and axis symmetry, we neglected the angular momentum flux taken away by radiation and the radial flux of radiation, we assumed large optical depth, used the vertical equilibrium formula derived by \cite{vertical} and we allowed for advective flux of energy. Slim disk equations were reduced to a two dimensional two boundary value problem. We solved it applying relaxation method \citep{numericalrecipes}. The solutions are presented in details in \S\ref{s.results}. For moderate and high mass accretion rates we observe large amount of advection which significantly changes the emergent flux profiles. For some particular range of accretion rates non-monotonic features in disk structure appear. We conclude they could be even more profound under different assumptions about disk vertical structure. The applications of the relativistic slim disk solutions presented here are discussed in \S\ref{s.prospects}.


\acknowledgements{The research reported here is a part of my doctoral
thesis work at the Copernicus Center in Warsaw,
Poland. I thank Marek Abramowicz, my thesis
supervisor, for suggesting the subject, many
discussions and his constant support. I also thank
Wlodek Klu{\'z}niak for his advice and help. I was
working on the problem at the Copernicus Center
and several other institutions: G\"oteborg University
(Sweden), Harvard-Smithsonian Center for Astrophysics (USA), Xiamen University
(China), Institute of Astronomy (Prague, the Czech
Republic) and at Nordita (Stockholm, Sweden).
I thank all of these institutions for hospitality
and support. This work was directly supported by
the Polish "Ph.D." grant N N203 304035 and also
by the grant N203 009 31/1466.
}

\bibliographystyle{apj}
\bibliography{ms}

\begin{appendix}
 \onecolumn
\section{The manual for tabulated slim disk solutions available online at\\ \texttt{http://users.camk.edu.pl/as/slimdisk}}
The solutions of the relativistic slim accretion disk model presented and discussed in this paper have been available online at \texttt{http://users.camk.edu.pl/as/slimdisk} since October 2009. In this section we present a brief description of the format they are provided in as well as for the included C interpolation routines.

There are two sets of files available: one for $10\msun$ BH only (weighting ca. 170 MB) and the other for a few BH masses with mass interpolation included (ca. 1 GB). The former is enough for ray-tracing purposes where the effective temperature and disk thickness profiles are the only needed while the latter for more general applications where other quantities, depending on BH mass in more complicated ways, are involved.

The disk profiles are provided in the following subfolders: \texttt{data-dsi/} in case of the single mass set of solutions and \texttt{data-dsi-mN/} in case of the full set of disk models where \texttt{N} stands for different BH masses and can take the following values: \texttt{5}, \texttt{10}, \texttt{20}, \texttt{30}, \texttt{50} and \texttt{100}$\msun$. Each folder contains a number of soltt.N.M.dat files describing accretion disks for given set of the input parameters. These files are indexed in the res.mamdot.dat file containing 8 columns with the following meaning:
\begin{enumerate}
 \item BH mass [$\msun$],
\item disk luminosity defined as: $\int_{r_{h}}^\infty 2\pi rF^{em}\rm dr$ in units of the Eddington luminosity:
$L_{Edd}=\frac{4\pi GMc}{\kappa_{es}}=1.25\times 10^{38} \frac M\msun \quad\rm erg\cdot s^{-1}$,
\item mass accretion rate in units of the critical mass accretion rate defined in Eq.~\ref{res.mdotc},
\item dimensionless spin parameter $a^*=\frac J{GM^2/c}$,
\item location of the sonic point in units of $2M$,
\item value of the angular momentum at the horizon (the eigen value of the transsonic solution) in units of $2M$,
\item \texttt{N} file index,
\item \texttt{M} file index.
\end{enumerate}
Each \texttt{soltt.N.M.dat} file contains 12 columns in the following order. All values but radius and temperature are given in $G=c=1$ units (conversion factors are given in Tab.~\ref{tab.conversion}).

\begin{enumerate}
 \item radius [$2M$],
\item radial velocity (as defined in Sect.~\ref{eq.equations}),
\item central temperature [K],
\item angular momentum,
\item surface density,
\item $H/r$ ratio,
\item vertically integrated pressure, 
\item radiation to gas pressure ratio,
\item flux emitted (both sides), 
\item flux advected, 
\item specific angular momentum ($u_\phi/u_t$),
\item undefined.
\end{enumerate}
The emitted flux $F^{em}$ is related to the disk effective temperature $T_{eff}$ by the following formula: $F^{em}=2\sigma T^4_{eff}$ where $\sigma=1.56\cdot10^{-60}\ \rm m^{-2}K^{-1}$ is the Stefan-Boltzmann constant in $G=c=1$ units.

\begin{table}[!h!c]
\caption{\textbf{Conversion factors from cgs to geometrical units}}
\vspace{2mm}
\begin{tabular}{l|l}

velocity                   &$1\ {\rm cm/s}=3.34\cdot 10^{-11}$\\\hline
angular momentum           &$1\ {\rm g\ cm^2/s}=2.48\cdot 10^{-43}\rm\ m^2$ \\\hline
surface density            &$1\ {\rm g/cm^2}=7.42\cdot 10^{-27}\rm\ 1/m$\\\hline
vertically integrated pressure & $1\ {\rm Ba\ cm}=8.26\cdot 10^{-48}\rm\ 1/m$\\\hline
flux                       &$1\ {\rm erg/cm^2/s}=2.76\cdot 10^{-56}\rm\ 1/m^2$\\
\end{tabular}\label{tab.conversion}
\end{table}

Disk solutions can be accessed directly or through provided interpolation routines. Two C files are attached to the archives: \texttt{dsi.c} and \texttt{draw\_dsi.c}. The former contains the interpolation routines and is designed to be easily included in any C code requiring instant access to disk solutions at any combination of the input parameters (mass accretion rate or luminosity, BH spin, radius and BH mass). The latter extracts given disk profiles for chosen input parameters, prints them to files and draw the solutions to \texttt{.GIF} and PostScript files using GnuPlot. Compilation of these routines requires GNU Scientific Library (\texttt{http://www.gnu.org/software/gsl}). The header file \texttt{dsi.h} as well as an exemplary \texttt{Makefile} are also included.

The interpolation routines take the same input parameters for both archives. The only difference is that the routines included in the smaller archive neglect the mass and always returns solutions for $10\msun$.

To include slim disk interpolation into a C code one should do the following:
\begin{enumerate}
 \item include \texttt{dsi.h} header file,
\item set up the interpolation by calling \texttt{dsi\_setup(MdotLum, a, M, alpha, MdotLumFlag, \&rmin, \&rmax)} with the following arguments:
\begin{itemize}
 \item \texttt{MdotLum} - mass accretion rate or disk luminosity in Eddington units depending on MdotLumFlag,
\item \texttt{a} - dimensionless spin parameter,
\item \texttt{M} - BH mass [$\msun$],
\item \texttt{alpha} - neglected,
\item \texttt{MdotLumFlag} - 0 for disk luminosity as the 1st argument, 1 for mass accretion rate,
\item \texttt{rmin} - minimal radius of the tabulated solution is returned,
\item \texttt{rmax} - maximal radius of the tabulated solution.
\end{itemize}
The routine returns 0 if the input parameters lie inside the grid of tabulated solutions and 1 if go outside the grid at least in one dimension.

\item call \texttt{dsi\_eval(r, N)} to evaluate a given parameter at radius \texttt{r} (in units of $2M$). The meaning of \texttt{N} is the following:\\ \\
\begin{tabular}{c|l}
 0 & absolute value of the radial velocity\\
 1 & central temperature\\
2 & specific angular momentum\\
3 & surface density\\
4 & H/r ratio\\
5 & total vertically integrated pressure\\
6 & radiation to gas pressure ratio\\
7 & flux emitted\\
8 & flux advected\\
9 & angular momentum\\
\end{tabular} \\ \\
All values are returned in geometrical units as discussed above.
\item set off interpolation for given disk parameters by calling \texttt{dsi\_setoff()}.
\end{enumerate}

The \texttt{draw\_dsi.c} file contains additional stand alone code which uses the routines provided by \texttt{dsi.c} to print out and draw disk solutions for given system parameters. The code takes input in the following format:\\
\hspace{.1\textwidth}\texttt{./draw\_dsi MBH N ml \%f a \%f [a \%f ml \%f] [a/ml \%f] filename}\\
where the meaning of the arguments is following:
\begin{itemize}
\item 
\texttt{MBH} - BH mass [$\msun$], 
\item \texttt{N} - index of the required quantity (as given above), 
\item \texttt{ml} - disk luminosity (if negative) or mass accretion rate (if positive), 
\item \texttt{a} - dimensionless BH spin, 
\item \texttt{filename} - name of output file for .gif and .ps figures. 
\end{itemize}
Examples:\\
\texttt{./draw\_dsi.out 10 3 ml .3 a 0 a .6 a .9 fig1}\\
\indent- plots surface density for M=$10\msun$ and the following (mdot, a) sets: (.3, 0), (.3, .6), (.3, .9)\\
\texttt{./draw\_dsi.out 10 7 a 0 ml -.1 ml -.2 ml -.3 fig1}\\
\indent- plots flux for M=$10\msun$ and the following (luminosity, a) sets: (.1, 0), (.2, 0), (.3, 0)\\
\texttt{./draw\_dsi.out 10 7 a 0 ml .6 a .9 ml .3 a .99 ml .1 fig1}\\
\indent- plots flux for M=$10\msun$ and the following (mdot, a) sets: (.6, 0), (.3, .9), (.1, .99)\\

\noindent Profiles of the required quantity for each set of disk parameters are printed out to \texttt{out\_n.dat} files each containing two columns with radius in the first one and profile of the required value in the second column.

The slim disk solutions are tabulated for $\alpha=0.1$, $0\leq a^*\leq0.99$ and $\mdot<500\mdot_{Edd}$.

The author will appreciate any comments and bug reports sent to \texttt{as@camk.edu.pl}.

\end{appendix}

\end{document}